# A model for the study of the Shubnikov-de Haas and the integer quantum Hall effects in a two dimensional electronic system

M. A. Hidalgo[(1)] and R. Cangas[(2)]

[(1)] *Departamento de Física, Universidad de Alcalá, Alcalá de Henares (Madrid), Spain*

[(2)] *Departamento de Física, Escuelta Técnica de Ingeniería Industrial, Universidad Politécnica de Madrid, Spain*

Electronic mail: miguel.hidalgo@uah.es, roberto.cangas@upm.es

## Abstract

Up to know all the experimental results concerning the integer and fractional quantum Hall effect are related to semiconductor heterostructures (and more recently with graphene). The common characteristic of all these systems is the presence of a reservoir of electrons, which, in fact, in the initial stage is the source of the electrons, providing the two-dimensional electron gas (2DES). Then, any physical realization of a 2DES is necessarily embedded in a 3D structure, which establishes the Fermi level. Hence, the 2DES appears to be an open system. In this paper we present an analytical approach to the integer quantum Hall effect (IQHE) and the Shubnikov-de Haas (SdH) phenomena in the 2DES, basing us in fundamental principles and showing the secondary role of the localized electron states in both phenomena. In fact, we show that the IQHE is a consequence of the fluctuations of electrons in the 2DES. Once we obtain the density of states of the 2DES under the application of a magnetic field we calculate both magnetoconductivities (diagonal and Hall) deducing them from the Boltzman semiclassical equation. The model proposed reproduces both phenomena, the width of the Hall plateaus (with the precision reached in the experimental measurements, of the order of $10^{-8}$-$10^{-9}$) and the corresponding minima of the diagonal magnetoresistivity, and also the dependence with temperature of the IQHE and SdH.

The quantum Hall effect is one of the most amazing and interesting phenomenon in the condensed matter physics discovered in the last part of the past century. It is characterized by the appearance of vanishing values in the diagonal conductivity (resistivity), $\sigma_{xx}$ ($\rho_{xx}$), at intervals of the magnetic field or the gate voltage for which the non-diagonal conductivity (resistivity), $\sigma_{xy}$ ($\rho_{xy}$), presents quantized plateaux in multiples of $e^2/h$ ($h$ the Planck constant and $e$ the fundamental charge) [1].

From the theoretical point of view, several attempts to understand the IQHE have been published. The most accepted one is based in the 'gendanken' experiment thought up by Laughlin, [2,3], explanation in which the SdH effect is not included. In his scenario, the localized states due to the ionized impurities and defects play a crucial role in the appearance of the plateaux [3]. However, this fact is against the experimental evidence, i.e., a bigger precision in the integer (and fractional) plateaux as a higher electron mobility of the electrons of the 2DES [4]. In fact, in view of the experiments the impurities and defects make the IQHE to be disappeared due to their effect over the Landau levels (see below). In spite of this problem, it is tacitly assumed that the IQHE is already explained. However, another problems remain also unexplained (the effect of the current intensity over both phenomena, IQHE and SdH, the asymmetry of the Landau levels due to the effect of the impurities…) [5,6].

More recently, the interest on these phenomena has been reactivated thanks to the observation of similar effects in graphene, a monolayer of graphite [7].

In this paper we present a model which allows to analytically study the IQHE and SdH, both together, in the whole magnetic field range and for both kind of experiments: $\rho_{xx}$, $\rho_{xy}$ *vs* $B$ and *vs* $V_g$.

We analyze the phenomena from a different point of view from Laughlin's theory, and basing us in fundamental principles, trying to understand those in terms of the fluctuation of the electron density. Any physical realization of a 2DES implies its open system character and, then, a fixed Fermi level determined by the whole structure where the 2DES is embedded [8]. This allows us to give an alternative view, similar to the initially tried at the beginning after the discovering of the IQHE, [9-13]. All of them considered a mechanism of pinning of electrons in the impurities as the origin of their fluctuation. However, this mechanism appears to be against the fundamental character of the phenomena.

Although theoretically one determines the magnetoconductivity tensor, experimentally are the resistance tensor (and, then, the resistivity one) the magnitudes measured. The relationships between the components of both tensors for a 2DES with a width $w$ and a length $l$ are given by $R_{xx} = l\rho_{xx}/w = l\left(\sigma_{xx}/\left[\sigma_{xx}^2 + \sigma_{xy}^2\right]\right)/w$ for the diagonal and $R_{xy} = \rho_{xy} = -\sigma_{xy}/\left[\sigma_{xx}^2 + \sigma_{xy}^2\right]$ for the Hall resistance, respectively.

The general Hamiltonian of an electron in a 2DES under the presence of a magnetic field is

$$H = \frac{1}{2m_e}(\vec{p} + e\vec{A})^2 + V(r) + H_s + U(r) + H_{deff} + H_{e-e} \tag{1}$$

where $\vec{A}$ is the magnetic potential vector and $(\vec{p} + e\vec{A})^2/2m_e$ the kinetic energy term, $m_e$ the free electron mass; $V(r)$ corresponds to the interaction of the electron with the crystal periodic potential; and $H_s$ the interaction of its spin with the magnetic field and the internal electron field (Zeeman and spin-orbit effects). The other terms to be considered are the interaction among electrons, $H_{e-e}$, with the ionized impurities, $U(r)$, and with the structural defects $H_{deff}$.

Assuming the effective mass approximation, we can rewrite the two first terms in the following way

$$H = \frac{1}{2m^*}\left(\vec{p} + e\vec{A}\right)^2 \tag{2}$$

where $m^*$ is the effective mass of the electron. The energy states of (2) correspond to the Landau levels

$$E_n = \left(n + \frac{1}{2}\right)\hbar\omega_0 = \left(2n+1\right)E_0 \tag{3}$$

where $n$=0, 1, 2,..., and $E_0 = \hbar\omega_0/2$ is the energy of the first Landau level with $\omega_0 = eB/m^*$.

Because its importance the next term to be considered is the interaction of the spin with the magnetic field and the internal electric field, i.e., the spin and spin-orbit interaction, $H_s$, which contributes to the energy state with a term $\pm g^* e\hbar B/4m_e = \pm g^* m^* \hbar\omega_c/4m_e$, and provides a splitting in every Landau level. $g^*$ is the generalized gyromagnetic factor, [14-17].

We will explain later the perturbative effect of the terms $U(r)$ and $H_{def}$ over the Landau levels; they will break smoothly the Landau degeneration. Finally, we assume that the term $H_{e-e}$ plays its role in the scattering time, (see below).

The density of the electron energy states in two dimensions without applying magnetic field is given by the expression $g_{B=0}\left(E\right) = g_0 = m^*/2\pi\hbar^2$, i.e., the states are uniformly distributed on energies (in this equation we have not taken into account the spin degeneration).

But when a magnetic field is applied the electron energy states change, appearing the Landau levels, degenerated in the values of the angular momentum. Then, every Landau level can be written as

$$g(x) = \int_{-\infty}^{\infty} g_0 \delta(z)\delta(z-x)\,dz \qquad (4)$$

where $x = E/\hbar\omega_c \mp g^* m^*/4m$, $\delta$ is the Dirac delta function and $g_0$ the density of states at zero magnetic field. Because the periodic character of the distribution of the Landau levels respecting to the index $n$, using the Poisson sum, we can write the whole Landau levels as

$$\sum_{n=-\infty}^{\infty} g\left(n+\frac{1}{2}\right) = \int_{-\infty}^{\infty} g(x)\,dx + 2\,\mathrm{Re}\left\{\sum_{p=1}^{\infty}(-1)^p \int_{-\infty}^{\infty} g(x)\,exp(2\pi x\, p\, i)\,dx\right\} \qquad (5)$$

Or in more useful way by

$$g(E) = g_0\left\{1+2\sum_{p=1}^{\infty}(-1)^p\, cos\left(2\pi p\left(\frac{E}{\hbar\omega_c} \mp \frac{g^* m^*}{4m}\right)\right)\right\} = g_0\left\{1+2\sum_{p=1}^{\infty} A_{s,p}\, cos(X)\right\}$$

with $X = 2\pi p\left(E/\hbar\omega_c - 1/2\right)$ and where the effect of the electron spin is included in the term:

$$A_{s,p} = cos\left(\pi p \frac{g^* m^*}{2m}\right)$$

In the generalized gyromagnetic factor is included the spin-orbit interaction due to the ionized impurities.

However, in a 2D real system there always are lattice defects and impurities, whose contribution to the density of states we assume to be to widen every Landau level. To include this effect in our model we have to impose such widening. In literature, there have been three main models: the semielliptical [18], lorentzian [19] and gaussian distribution functions, [8]. We will take the third. This implies that instead of (Eq. 4) we have to start from

$$g(x) = \int_{-\infty}^{\infty} g_0 \sqrt{2\pi}\,\frac{\hbar\omega_c}{\Gamma}\, exp\left(-\frac{2\pi^2 z^2}{\left(\Gamma/\hbar\omega_c\right)^2}\right)\delta(z-x)\,dz \qquad (6)$$

Where $\Gamma$ is the width of the gaussian function, the parameter directly related to the effects of impurities and defects over the 2DES. In the simulations we present below we will assume it to be constant as a function of the magnetic field as a first approach.

Then, using again the Poisson sum we obtain now

$$g(E) = g_0 \left\{ 1 + 2\sum_{p=1}^{\infty} A_{S,p} A_{\Gamma,p} cos\left[ 2\pi p \left( \frac{E}{\hbar\omega_c} - \frac{1}{2} \right) \right] \right\} \tag{7}$$

Hence, the effect of impurities and defects is included in the analytical model in the term

$$A_{\Gamma,p} = exp\left[ -\frac{1}{2}\left( \frac{p\Gamma}{\hbar\omega_c} \right)^2 \right]$$

A physical realization of a 2DES is obtained confining electrons in the inversion layer of a semiconductor heterostructure, such as AlGaAs-GaAs, or in a silicon MOSFET (Metal Oxide Semiconductor Field Effect Transistor) device and, more recently, in a graphene layer. Then, the main point of our view is to consider a Fermi level always matched by the whole 3D system where the 2DES is embedded.

In all the discussion presented below, we will assume low temperatures and low transport electric fields.

Once we have the density of states of the 2DES, we can deduce the expressions for both magnetoconductivities, Hall and diagonal (SdH). When an electric field $\vec{E}^t = \left(E_x^t, E_y^t, 0\right)$ and a magnetic field $\vec{B} = \left(0, 0, B\right)$ are applied to the 2DES (places in the *x-y* plane), the velocity of electrons of the gas can be decomposed in two components, $\vec{v} = \vec{v}_d + \vec{v}_c$, the first term corresponding to the drift velocity, $\vec{v}_d = \left(\vec{E}^t \times \vec{B}\right) \big/ B^2 = \left(E_y^t, -E_x^t, 0\right) \big/ B$, important at low magnetic fields, and the second to the quantized cyclotron velocity. We assume that the energy states are completely determined by the cyclotron energy and, then, the wave vector $\vec{k}_c = m^* \vec{v}_c / \hbar$ is a good quantum number to describe the electron states at high magnetic fields, i.e., $\omega_c \tau > 1$, where $\tau$ is the lifetime of the corresponding state (see below).

The general expression for the electron current density is

$$\vec{j} = \frac{e}{4\pi^2} \int_{\vec{k}_c} \vec{v} f\left(\vec{k}_c\right) d^2\vec{k}_c = e \int \vec{v} f\left(E\right) g\left(E\right) dE \tag{8}$$

Where $f\left(\vec{k}_c\right) = f^0\left(\vec{k}_c\right) + f^1\left(\vec{k}_c\right)$ is the distribution function under the application of the fields, $f^0$ the Fermi-Dirac equilibrium one and $f^1$ the non-equilibrium supplement to be defined, and due to the magnetic field and the electric field $E^t$. We have to suppose that $f^1 << f^0$.

At high magnetic fields, the integration of Eq. (8) is over all possible states of the electron. Thus, we have

$$\vec{j} = \frac{e}{4\pi^2} \int_{\vec{k}_c} \vec{v} \left( f^0\left(\vec{k}_c\right) + f^1\left(\vec{k}_c\right) \right) d^2\vec{k}_c = \vec{j}^0 + \vec{j}^1$$

The first term is

$$\vec{j}^0 = \frac{e}{4\pi^2} \int_{\vec{k}_c} \vec{v} f^0\left(\vec{k}_c\right) d^2\vec{k}_c = e\vec{v}_d \int_{\vec{k}_c} f^0\left(\vec{k}_c\right) d^2\vec{k}_c = \frac{en}{B}\left(E_y^t, -E_x^t\right) \tag{9}$$

Where $n$ is the electron density of the 2DES. Hence, $\vec{j}^0$ corresponds to the Hall current density, and the corresponding magnetoconductivity at high magnetic fields will be

$$\sigma_{xy} = -\sigma_{xy} = -\frac{en}{B} \tag{10}$$

What implies that the Hall magnetoconductivity at high magnetic field is an equilibrium property of the 2DES. The electron density is determined through the equation

$$n = \frac{1}{2\pi^2} \int_{\vec{k}_c} f^0\left(\vec{k}_c\right) d^2\vec{k}_c = \int_0^{\infty} f^0\left(E\right) g\left(E\right) dE \tag{11}$$

At low temperatures, $E_F >> kT$, and using the Poisson sum (Eq.7) yields, [8]

$$n = n_0 + \frac{2eB}{h}\sum_{p=1}^{\infty}\frac{1}{\pi p}A_{S,p}\ A_{\Gamma,p}\ A_{T,p}\ sin\left[X_F\right] = n_0 + \delta n \tag{12}$$

Where the first term the electron density of the 2DES at zero magnetic field, $n_0$; and

$$A_{T,p} = \frac{z}{sinh\ z}$$

With $z = 2\pi^2 pkT/\hbar\omega_c$ and $X_F = 2\pi p\left(E_F/\hbar\omega_c - 1/2\right)$. This factor takes into account the effect of the temperature over the gas. The second term of (Eq.12), $\delta n$, takes into account the fluctuations of the electron density and is a direct consequence of the fixed Fermi level by the whole system. As we mention before, the semiconductor surrounding of a 2DES acts as drain and source of electrons. When a Landau level crosses the Fermi level a charge transfer takes place from the 2DES and the reservoir of impurities; and the charge flux is in the contrary sense when the Fermi level is in between two Landau levels, due to the increase in the density of states as a consequence of the increasing magnetic field.

Consider now the second term of the current density, $\vec{j}^1$. At high magnetic fields we have

$$\vec{j}^1 = \frac{e}{4\pi^2}\int_{\vec{k}_c}\vec{v}f^1\left(\vec{k}_c\right)d^2\vec{k}_c = \frac{e}{4\pi^2}\left[\int_{\vec{k}_c}\vec{v}_c f^1\left(\vec{k}_c\right)d^2\vec{k}_c + \int_{\vec{k}_c}\vec{v}_d f^1\left(\vec{k}_c\right)d^2\vec{k}_c\right] \tag{13}$$

Then we need to know $f^1\left(\vec{k}_c\right)$. For this task we have to use the Boltzman equation, linearized in the electric field, i.e.,

$$-\frac{e}{\hbar}\left(\vec{E}^t + \vec{v}\times\vec{B}\right)\cdot\frac{\partial f}{\partial\vec{k}_c} = -e\left(\vec{v}_c\cdot\vec{E}^t\right)\frac{\partial f^0}{\partial E} - \frac{1}{\hbar}\left(\vec{v}_c\times\vec{B}\right)\cdot\frac{\partial f^1}{\partial\vec{k}_c} = -\left(\frac{\partial f}{\partial t}\right)_{collision} \tag{14}$$

And to determine the collisions term we will assume very low temperatures and an electric field $\vec{E}^t$ not too strong (this fact allows to vanish quadratics terms in the electric field which corresponds to deviations of the Ohm law). Using the time dependent perturbation theory we can obtain a similar expression than the obtained at low magnetic fields, $\left(\partial f/\partial t\right)_{collision} = f^1/e\tau$, where $1/\tau\left(E\right) \approx \int_{E'\in D(E')} \varpi\left(E,E',t\right) g\left(E'\right) dE'$, and $\tau\left(E\right)$ the lifetime of each Landau level, related to the probability by time unit that a transition takes place between the states, $\varpi\left(E,E',t\right)$. The lifetime has not to be the same for the different states of a Landau level. Then, this term can explain the observed behaviour of the SdH effect with the current intensity [5,6].

Looking for solutions like $f^1 = -e\tau\left(-\partial f^0/\partial E\right)\left(\vec{v}_c \cdot \vec{A}\right)$, we obtain $\vec{A} = \alpha\left(\vec{E}^t + e\tau\left[\vec{B}\times\vec{E}^t\right]/m^*\right)$ where $\alpha = 1/\left(1+\omega_c^2\tau^2\right)$, [20]. The term associated with the drift velocity $\vec{v}_d$ of (Eq.13) is vanished for being of the order of $\vec{E}^{t2}$, i.e., terms out of the Ohms law. After some algebraically calculations we can get

$$\vec{j}^1 = \frac{e^2}{4\pi^2}\int_{\vec{k}_c} \tau\alpha\left(-\frac{\partial f^0}{\partial E}\right)\left(v_x^2 E_x^t, v_y^2 E_y^t\right) d^2\vec{k}_c \qquad (15)$$

Where we have written $\vec{v}_c = v_x\vec{u}_x + v_y\vec{u}_y$, in the 2D plane. This leads to the diagonal conductivity at high magnetic fields

$$\sigma_{ii} \approx \frac{e}{\omega_c B}\int_{E=0}^{\infty}\left(-\frac{\partial f^0}{\partial E}\right) E g\left(E\right) dE \int_{\eta\in D(E')} g\left(E+\eta\right)\varpi\left(\eta,t\right) d\eta \qquad (16)$$

With the index $i=x, y$ and $\eta = E' - E$. Hence, we can associate the term $\vec{j}^1$ of the current density with the diagonal magnetoconductivity. Assuming isotropy in the S2D we find $\sigma_{xx} = \sigma_{yy}$.

If we assume energy conservation in the collision term, (Eq.16) is simplify to

$$\sigma_{ii} = eN/B\omega_c\tau \qquad (17)$$

Where $N$ is the electron density at the Fermi level, (and at zero temperature). Because $\partial f^0(E)/\partial E = -\delta(E - E_F)$, $N = E_F g(E_F)$, being $g(E_F)$ the density of states at the Fermi level. If we take into account that $E_F = \hbar^2 n_0 \pi / m^* = n_0/g_0$, then we get

$N = n_0 g(E_F)/g_0$

We have selected two of the good reference measurements in literature of both phenomena, the SdH and IQHE effects as a function of the magnetic field, [21,22], in order to simulate them with the model and to check it. In both cases, we have considered Gaussian shape for the Landau levels to take into account the effect of the impurities over the electrons of the gas.

In Figure (1) we present the simulation of both phenomena for a sample in the conditions of reference [21], with a electron density at zero magnetic field $n_0 = 3.7\times10^{15} m^{-2}$ and a mobility $\mu = 4.1\times0^4\, cm^2/Vs$; and in Figure (2) for one of the samples analyzes in the reference [22], where the electron density at zero magnetic field is in this case $n_0 = 4\times10^{15} m^{-2}$ with a mobility $\mu = 10^5\, cm^2/Vs$.

Comparing with the experimental results the accordance of the model is quite good, being the only important difference in the amplitude of the SdH oscillations, determined in the simulations of the model by (Eq.17).

We have presented an approach to the IQHE and SdH based in first principles. Basing us in the developed model, both phenomena appear to be a consequence of the electron fluctuation in the 2DES under the application of a magnetic field, due to the open character of any physical realization of a two-dimensional electron gas.

In light of the results obtained for the IQHE and SdH, in a forthcoming works, we will present extensions of the presented approach to the fractional quantum Hall effect and the anomalous IQHE recently observed in graphene.

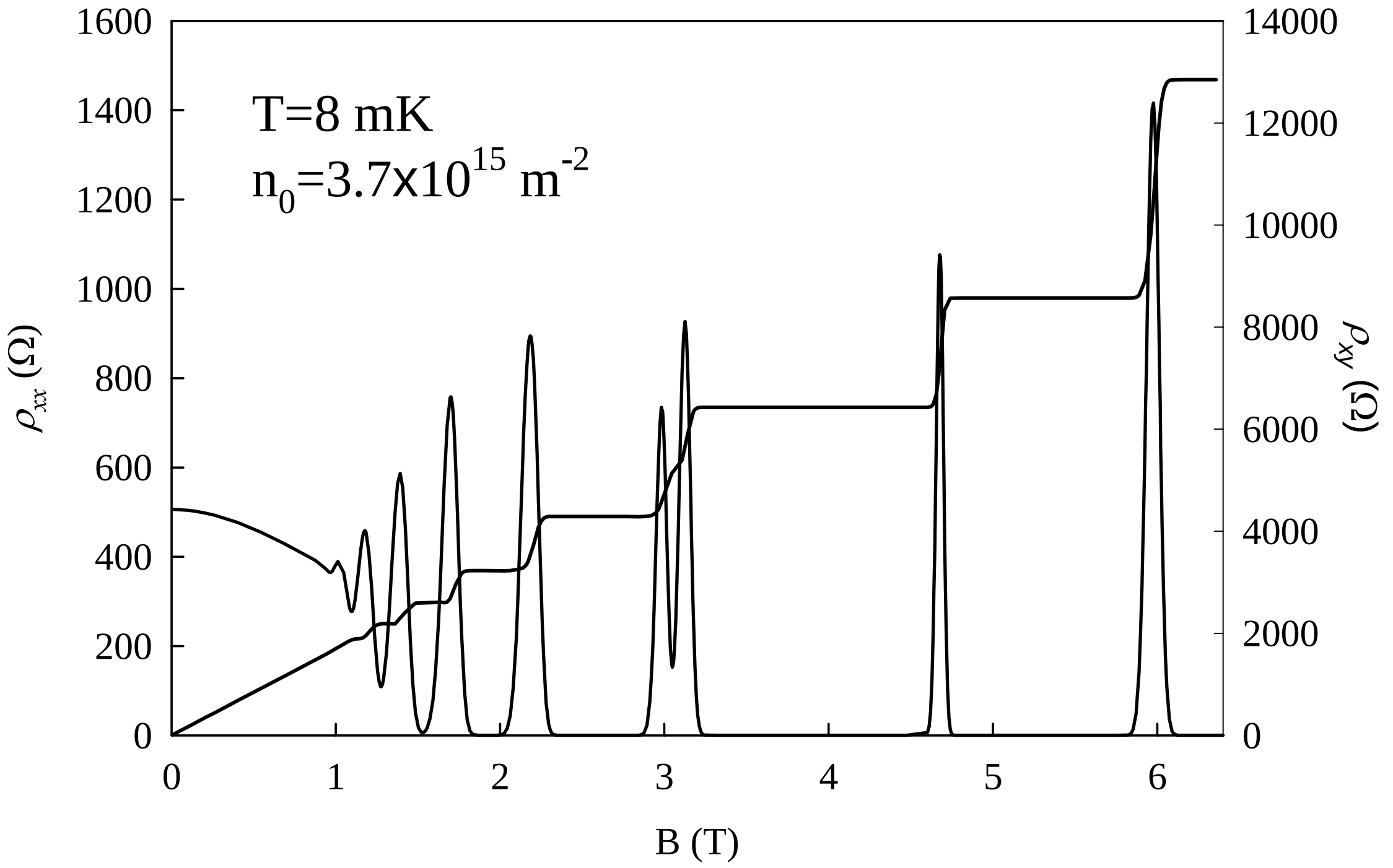

Figure 1: Simulation of the Shubnikov-de Haas and integer quantum Hall effects measurements as obtained by Klitzing, [21].

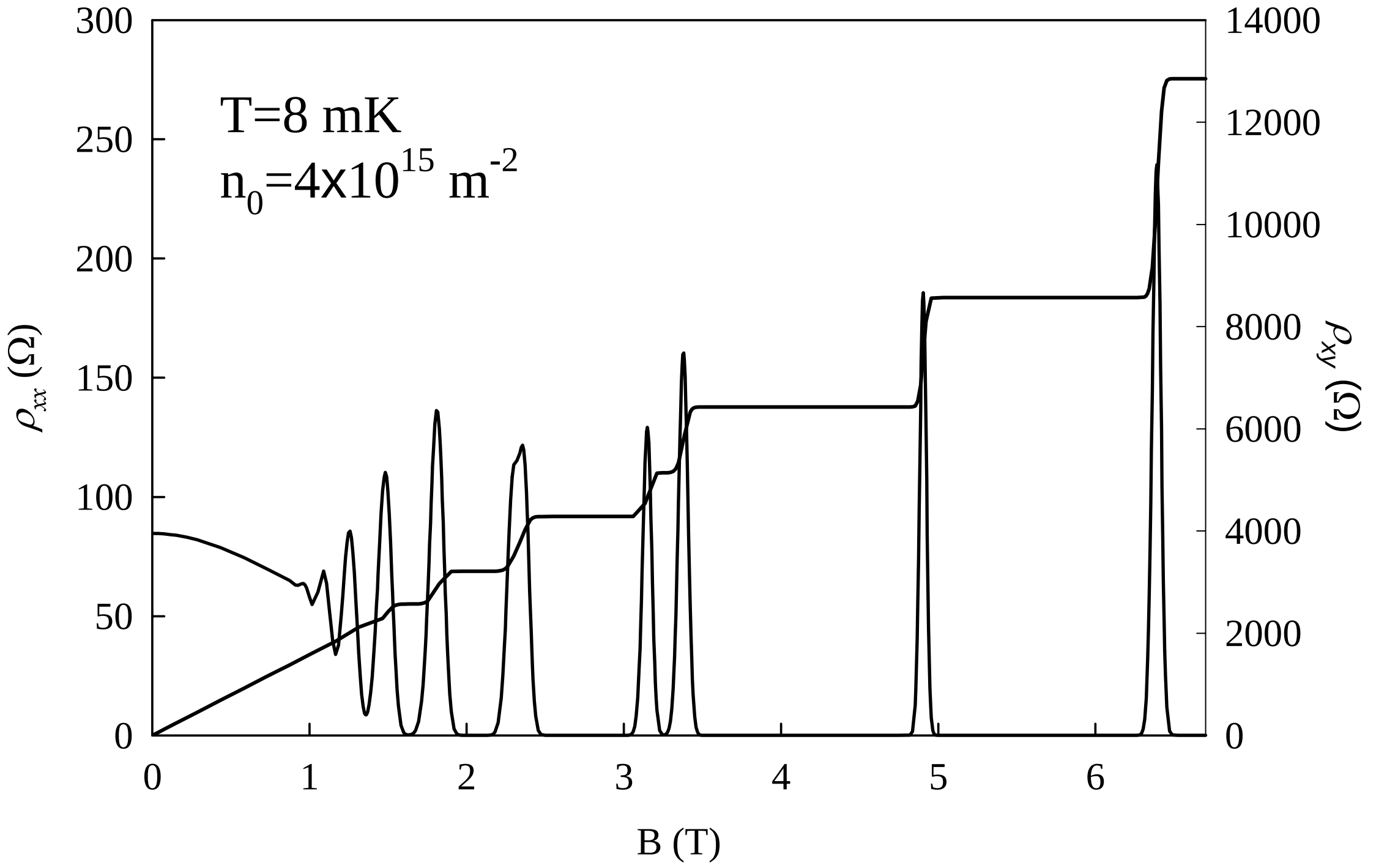

Figure 2: Simulation of the Shubnikov-de Haas and integer quantum Hall effects measurements as obtained by Ebert *et al.*, [22].